\documentclass[aps,prl,twocolumn]{revtex4}
\usepackage{graphicx}
\usepackage{psfrag}
\usepackage{amsmath}

\newcommand{\ba}{\begin{array}}
\newcommand{\ea}{\end{array}}
\newcommand{\bc}{\begin{center}}
\newcommand{\ec}{\end{center}}
\newcommand{\bee}{\begin{eqnarray}}
\newcommand{\eee}{\end{eqnarray}}
\newcommand{\br}{\mbox{\boldmath $r$}}

\newcommand{\m}{\mbox{\boldmath $\mu$}}
\newcommand{\bF}{\mbox{\boldmath $F$}}
\newcommand{\nnabla}{\mbox{\boldmath $\nabla$}}

\begin{document}

\title{%\Large 
Symmetric three-particle motion 
in Stokes flow: equilibrium for heavy spheres in contrast to 
``end-of-world'' for point forces}
\author{Maria L. Ekiel-Je\.zewska}
\affiliation{Institute of Fundamental Technological Research,
    Polish Academy of Sciences, \'Swi\c etokrzyska 21, 00-049 Warsaw, Poland}
\author{Eligiusz Wajnryb}
\affiliation{Institute of Fundamental Technological Research,
    Polish Academy of Sciences, \'Swi\c etokrzyska 21, 00-049 Warsaw, Poland}

\date{21.08.2005}

\begin{abstract}
A stationary stable solution of the Stokes equations
for three identical heavy solid spheres falling in a vertical plane is found.
It has no analog in the point-particle approximation. Three 
spheres aligned horizontally at equal distances evolve towards the equilibrium 
relative configuration 
while the point particles 
collapse onto a single point in a finite time.
\end{abstract}

\maketitle

In recent experiments, the structure and velocity field of a sedimenting 
non-Brownian suspension 
have been observed to differ significantly from the equilibrium 
state~\cite{Ramaswamy}. Theoretical explanation of this behavior 
must take into account hydrodynamic interactions between the suspended 
particles.
The simplest interesting cluster consists of three identical point particles 
settling under gravity  in a
quiescent viscous infinite fluid. Its evolution  was 
analyzed numerically and turned out to be sensitive to small changes of 
the initial configuration~\cite{janosi}. 
Symmetric periodic motions of three point-particles and of three spheres 
located at vertices of an isosceles triangle
have been also found~\cite{hocking,caflish} and the equilateral  horizontal
triangle is known as the equilibrium solution. 
Until now, theoretical results confirmed the analogy between the 
motion of three well-separated spheres and three point-particles.

In this work a simple example is studied, which 
illustrates that evolution of 
well-separated spheres may differ significantly from 
the corresponding point-particle approximation.
Settling of 
three identical particles 
under gravity is analyzed. The
particle  centers are located at vertices of an isosceles vertical 
triangle with 
the horizontal base.
First, the system of equations is specified, and the method 
to solve it is outlined. Then, 
a new equilibrium relative configuration of the spheres is found. Next, 
the particles are initially aligned horizontally and the 
motion of the sphere centers is determined and compared with the 
motion evaluated within the point-particle approximation.  
Finally, other relative trajectories of the spheres 
are determined and stability of the 
equilibrium is shown.

A low-Reynolds-number fluid flow is considered. Its velocity 
${\bf v}(\br)$ and pressure 
$p(\br)$ satisfy the Stokes equations~\cite{happel},
\bee
\eta {\bf \nnabla}^2 {\bf v} -{\bf \nnabla} p = {\bf 0}, &\hspace{2cm}& 
{\bf \nnabla} \cdot {\bf v} = 0,\label{incompressible}
\eee
with the fluid viscosity $\eta$. The fluid is infinite. Its motion is 
generated by settling 
of three identical spheres under gravity $\bF=-F \hat{\bf e}_z$. 
The stick boundary conditions are assumed at the surfaces of the spheres.
Positions of the sphere centers $\br_i(t)$ satisfy the following 
equations,
\bee
{\dot{\br}}_i(t)  &=& \left[ \sum_{k=1}^3\m_{ik}\right] \cdot  \bF, 
\hspace{1cm} i=1,2,3.\label{em}
\eee
The $3\times3$  mobility 
matrices $\m_{ik}$ depend on the instantaneous relative configuration 
of all the spheres, and are 
evaluated numerically by the multipole expansion  \cite{kim,fel}.  
The algorithm from 
Ref.~\cite{cfhwb} and its accurate numerical FORTRAN implementation 
described in Ref.~\cite{jcp} are applied, with 
the multipole order $L=4$. The set of the ordinary differential 
equations \eqref{em} is solved numerically by the adaptive fourth-order 
Runge-Kutta method~\cite{NR}. In the following, distances will be  
normalized by the sphere diameter $d$ and time by two Stokes times 
$\tau_s=3\pi\eta d^2/F$, keeping the same notation. The dimensionless 
variables satisfy Eq.~\eqref{em} with $F=1$.

Initially, the sphere centers are 
located at vertices of an isosceles vertical triangle with the horizontal 
base. From the symmetry of the Stokes equations it follows 
that the time-dependent configuration is also of this type.  
The apex sphere is labeled 2 and the two other spheres are 
labeled 1 and 3.
The relative positions $\br_{12}\equiv \br_1-\br_2$ and 
$\br_{32}\equiv \br_3-\br_2$ are parameterized as
\bee
\br_{12}&=&(-x/2,0,z),\\
\br_{32}&=&(x/2,0,z).
\eee
The distance $x$ between the twin spheres 1 and 3 and the vertical 
separation $z$ between the twins and the apex sphere 2 satisfy the 
following system of equations,
\bee
\dot{x} &=& v_x(x,z), \label{d1}\\
\dot{z} &=& v_z(x,z), \label{d2}
\eee
with the initial values $x(0)=x_0$, $z(0)=z_0$.
Here $v_x=\sum_{k=1}^3[\mu_{3k,xz}-\mu_{2k,xz}]$,  
and $v_z=\sum_{k=1}^3[\mu_{3k,zz}-\mu_{2k,zz}]$, with
the Cartesian components of $\m_{ik}$
dependent only on $x$ and $z$.
Once $x(t)$, $z(t)$ are evaluated, then  $\br_2=[0,0,z_2(t)]$
is obtained by a direct integration of Eq. \eqref{em}.

In the point-force approximation, the same units are used, and 
a single point moves with  the 
Stokes velocity of the single sphere. 
The point-particle dynamics reads,
\bee
\dot{\br}_i &=& - \sum_{k\ne i} {\bf T}_{ik}\cdot \hat{\bf e}_z - 
\hat{\bf e}_z,
\label{pp}
\eee
with the dimensionless Oseen 
tensor,
\bee
{\bf T}_{ik} &=& \frac{3}{8r_{ik}} 
({\bf I} + \hat{\br }_{ik} \hat{\br }_{ik}),
\eee
the unit vectors $\hat{\br }_{ik}=
\br_{ik}/r_{ik}$ and the unit tensor ${\bf I}$. Eqs.~\eqref{pp} are 
integrated numerically by the adaptive fourth-order Runge-Kutta 
method~\cite{NR}.

\begin{figure}[h]
\psfrag{v12}{\huge $\;\;\;\;\;\;v_z(1,z)$}
\psfrag{z12}{\huge $\!\!\!z$}
\resizebox{8.6cm}{!}{\includegraphics{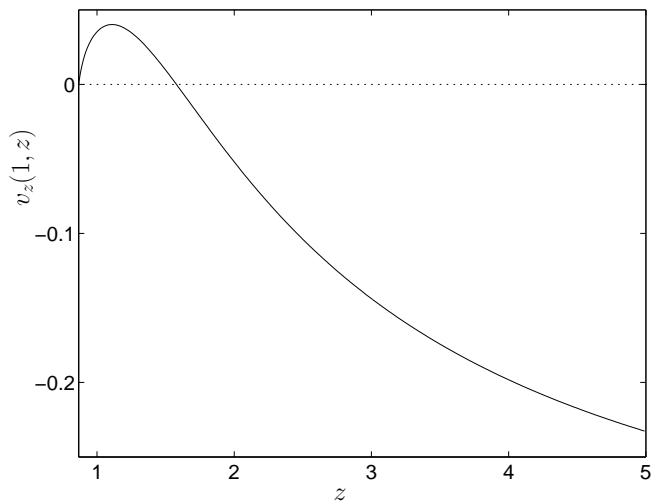}}
\caption{Motion of the 
touching spheres with respect to the single one.}\label{zlepek}
\end{figure}

Now, equilibria for the dynamical system 
\eqref{d1}-\eqref{d2} will be found.  
The r.h.s. of Eq.~\eqref{d1} vanishes if the twin spheres~1 and~3 touch 
each other and the distance between their centers
$x=1$. In this case, the 
lubrication forces~\cite{JO} prevent spheres 1 and 3  
from a relative motion. Therefore 
$v_x(1,z)=0$ at each point $z$, and the motion is vertical. At 
equlibrium point $\tilde{z}$, also 
$v_z(1,\tilde{z})$ should vanish.
Consider first configurations with  
the single sphere~2 below the touching dublet. In this case $z$ is positive;
moreover, $z\ge \sqrt{3}/2$, because the spheres do not overlap.
The function $v_z(1,z)$  is evaluated 
numerically and plotted in Fig.~\ref{zlepek}. 
Two equilibria are seen. The first one, with $(x,z)=(1,\sqrt{3}/2)$, 
corresponds to the touching
triplet of spheres, with relative motions excluded 
by the lubrication forces~\cite{JO}. 
It is interesting to see that there exists also another positive root 
$z_{eq}$ of the equation $v_z(1,z)=0$.
For a given multipole order $L$, $z_{eq}$ is evaluated
numerically by the standard bisection method~\cite{NR}. 
Next, $z_{eq}$ is evaluated for all $L=3,...,30$ and the limit 
$L\rightarrow \infty$ is taken,
\bee
z_{eq}=1.578634.
\eee

The question arises does this equilibrium attract trajectories. First, 
vertical motion of the horizontal touching doublet located above the 
singlet will be analyzed, 
using the function plotted in Fig.~\ref{zlepek}.
In the limit of infinite $z$, the faster 
doublet and 
the slower singlet are not influenced by each other and as a result 
$\dot{z}=-0.3799554$, 
see e.g.~\cite{free}. Starting from any $z_0>z_{eq}$, 
the relative 
distance $z$ between the doublet and the singlet decreases with time, 
because the vertical velocity $v_z$ is negative. However, if $z_0<z_{eq}$, 
then  $v_z>0$ and $z$ increases with time. It is remarkable 
that in this case a heavier doublet is repelled by a lighter singlet located 
below. In both cases, the system tends to the equilbrium 
position $z_{eq}$, approaching it after infinite time.  
Indeed, from 
Fig.~\ref{zlepek} it is clear that  
$\partial v_z(1,z)/\partial z<0$, if $z$ is close to $z_{eq}$,
and therefore $|z-z_{eq}|$ decreases with time exponentially. 

Consider now configurations 
where the touching spheres are 
below the single one, with $z<0$. The Stokes equations are invariant 
with respect to the time reversal, 
supperposed with the 
reflection in the horizontal plane, which contains the center of sphere~2.
Therefore 
\bee
v_x(x,z)\!&=\!&-v_x(x,-z),\hspace{0.7cm} %\label{sy1}\\
v_z(x,z)=v_z(x,-z).\hspace{0.7cm}\label{sy2}
\eee
In particular, for $z<0$ the relative velocity is immediately obtained from 
Fig.~\ref{zlepek}, using the relation $v_z(1,z)=v_z(1,-z)$. 
Starting from negative $z_0$, trajectories escape from $-z_{eq}$. 
If $z_0>-z_{eq}$, then the doublet is attracted by 
the singlet and $z\rightarrow -\sqrt{3}/2$. If $z_0<-z_{eq}$, then the 
doublet is repelled by the singlet and $z\rightarrow -\infty$.
To conclude, the dynamics $\dot{z}=v_z(1,z)$
has four equilibria, with $x=1$ and $z=\pm z_{eq}$ or 
$z=\pm\sqrt{3}/2$. 

\begin{figure}
[h]
\psfrag{xi}{\Large $x_i$}
\psfrag{zi}{\Large $z_i$}
\resizebox{8.6cm}{!}{\includegraphics{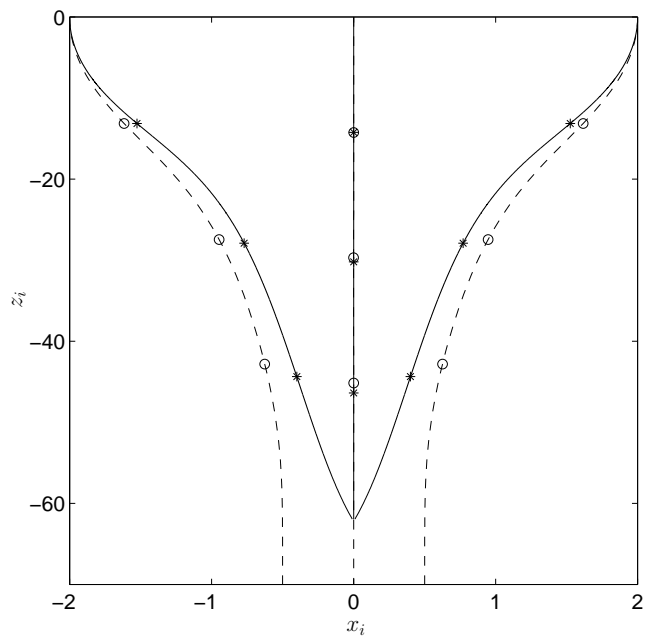}}
\caption{Trajectories of three sphere centers (dashed lines, circles) and  
three point-particles (solid lines, stars), at $t=0$ 
centered at $x_1=2=-x_3$, 
$x_2=0$, $z_1=z_2=z_3=0$. The symbols represent positions at times t=10, 
20 and 30.}\label{ps}
\end{figure}

Now, general case will be studied when spheres 1 and 3 do not touch each 
other. Assume first that initially three 
sphere centers are aligned horizontally. 
In Fig.~\ref{ps},
typical trajectories $z_i(x_i)$ of spheres and point-particles are 
compared. Here 
the solution of the 
dynamics \eqref{em} is denoted as $\br_i(t)=[x_i(t),0,z_i(t)]$, 
with $i=1,2,3$. 
At the beginning, trajectories of the point 
particles coincide with those of the spheres. However, after a finite time 
$t= 36.74$ they collapse onto a single point, and obviously the point-force 
approximation breaks down. 

\begin{figure}
[h]
\psfrag{ln xi}{\huge $\ln \xi$}
\psfrag{t}{\huge $\!\!\!t$}
\resizebox{8.6cm}{!}{\includegraphics{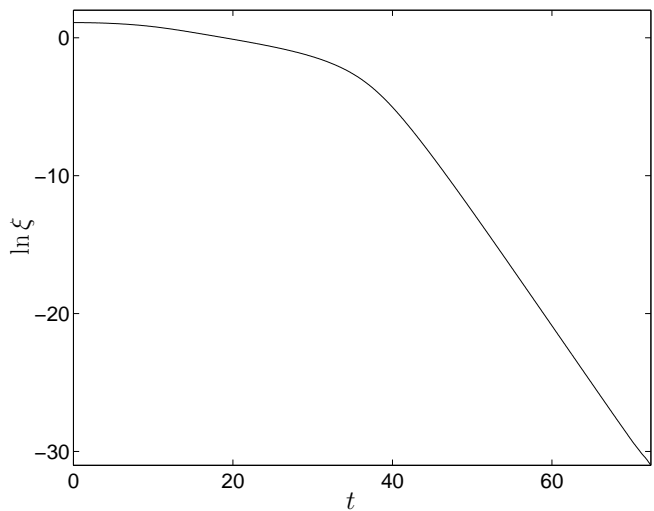}}
\caption{Time-dependent size $\xi$ of the gap between surfaces of the 
spheres 1 and 3; initially $\xi_0=3$ and $z_0=0$.} \label{xit}
\end{figure}

\begin{figure}[t]
\psfrag{z}{\huge $z$}
\psfrag{t}{\huge $\!\!\!t$}
\psfrag{zz}{\huge $\!\!z_{\!eq}$}
\psfrag{zzs}{\Huge $\!\!z_{\!eq}$}
\resizebox{8.6cm}{!}{\includegraphics{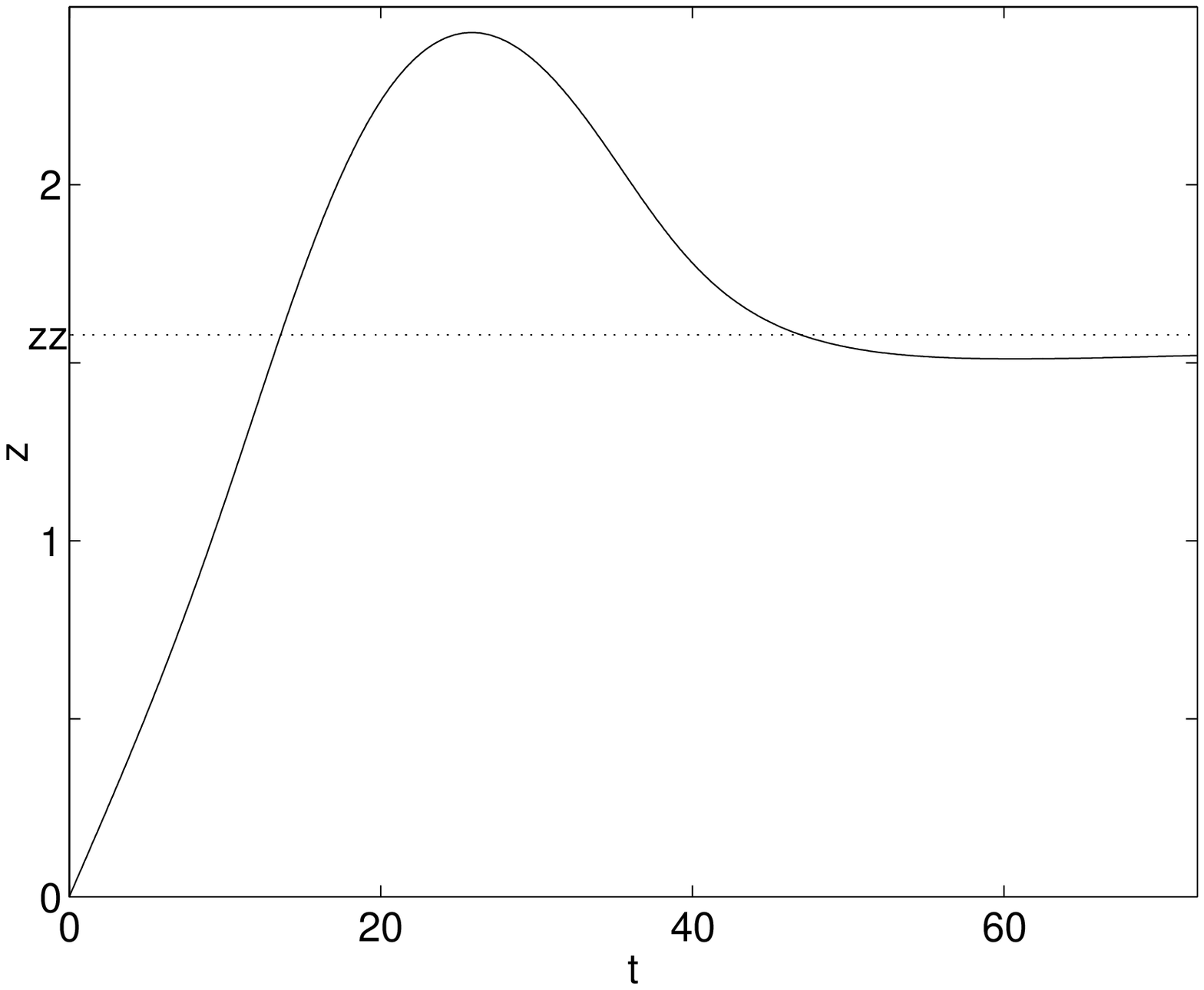}}
\begin{picture}(1,1)(0, -30.5)
\resizebox{4.1cm}{!}{\includegraphics{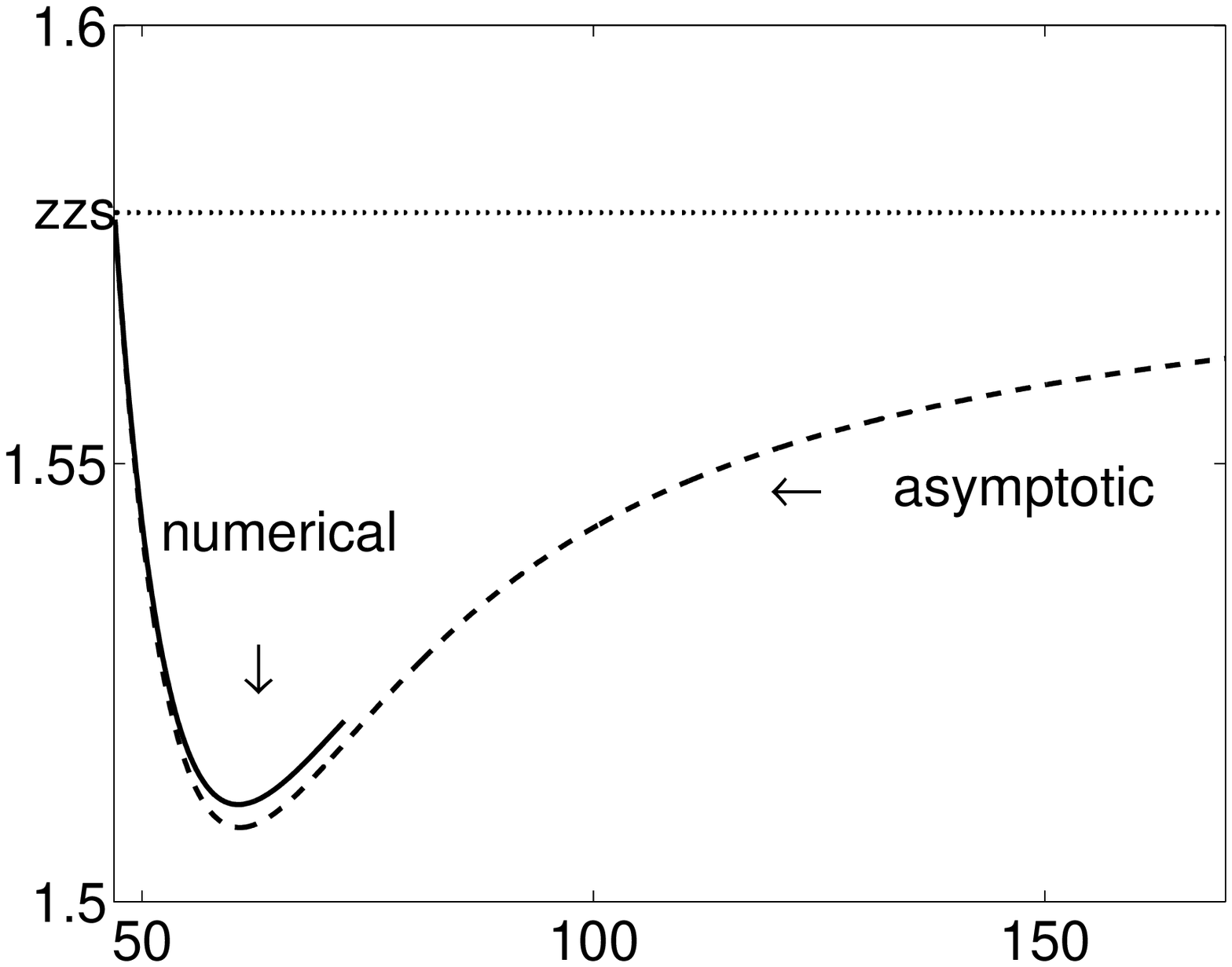}}
\end{picture}
\caption{Evolution of the vertical separation $z(t)$ from the twin spheres 
1,3 
to sphere 2  (numerical result).
Initially $x_0=4,$ $z_0=0$. Inset: comparison with the 
asymptotic expression.}\label{zt}
\end{figure}

In the following, only the relative two-dimensional dynamics 
\eqref{d1}-\eqref{d2} will be discussed. Evolution of  the initial values 
$x_0$ and $z_0=0$ is of special interest.
The distance $\xi=x-1$ between the surfaces of spheres 1 and 3 
decreases monotonically, decaying exponentially to zero 
for long times, as illustrated in Fig.~\ref{xit}. %
For close twin spheres, if 
$x_0$ is sufficiently small, then 
their vertical separation $z$ from sphere 2 
increases with time. For a large $x_0$, at the beginning 
$z$ increases, significantly exceeding $z_{eq}$, then drops down 
below $z_{eq}$ to approach 
it slowly again, as depicted in Fig.~\ref{zt}. 
Numerical 
integration of Eqs. \eqref{d1}-\eqref{d2} is limited to the gap sizes 
$\xi \lesssim 10^{-14}$. It is remarkable that for such close distances, the 
vertical separation $z$ still differs from its equilibrium value $z_{eq}$ by 
around 4\%. This difference is related to rotation of spheres 1 and 2, 
significant even for such tiny gaps between their surfaces.

Close to the equilibrium, for 
very small $\xi$ and small $z-z_{eq}$, it is possible to solve the 
dynamics~\eqref{d1}-\eqref{d2}, using the asymptotic lubrication 
expansion of the two-sphere mobilities \cite{JO} for
spheres 1 and 3. To extract the leading terms in the limit of $\xi 
\rightarrow 0$, the velocity 
of each sphere 1 and 3 is written as superposition of velocities 
calculated for two other 
problems in the absence of sphere 2:
sedimentation of the spheres 1 and 3 and 
the free motion of the spheres 1 and 3 in the ambient flow 
determined by the dynamics~\eqref{d1}-\eqref{d2}. On the other hand, the 
free motion is obtained by applying the two-sphere mobility to the forces 
${\bf F}_{0i}$ and torques ${\bf T}_{0i}$ exerted on spheres $i=1,3$ 
fixed in the same ambient flow. Therefore
\bee
\dot{x_1} &=& (x^A_{11}-x^A_{13}) F_{01x}, \\
\dot{z_1} &=& (y^A_{11}+y^A_{13}) (F_{01z}-1)+3(y^B_{11}+y^B_{13}) T_{01y},
\hspace{0.8cm}  
\eee
with the components of the two-sphere mobility denoted as 
in Ref.~\cite{JO} by $x^A_{11}$, $x^A_{13}$, $y^A_{11}$, $y^A_{13}$,  
$y^B_{11}$ and $y^B_{13}$. The force and torque units are $F$ and $Fd$, 
respectively. For sphere 3, the corresponding 
formulas follow from the symmetry with respect to reflections in the plane 
$x=0$. For $\xi \rightarrow 0$, the forces ${\bf F}_{01}={\bf F}_{03}$ and 
torques 
${\bf T}_{01}=-{\bf T}_{03}$, as well as  the velocity of sphere 2, are 
regular functions 
of $\xi$. From Ref.~\cite{JO} it follows that $(x^A_{11}-x^A_{13}) 
\sim \xi$, while $(y^A_{11}+y^A_{13}) \sim 1/(\ln \xi +B) + \mbox{const}$ 
and $(y^B_{11}+y^B_{13}) \sim 1/(\ln \xi +B)$, with $B=4.00$.
Therefore for very small $\xi$,  
\bee
v_x(x,z)&=&f(z)\,\xi +{\cal O}(\xi^2 \ln \xi),\label{scx}\\
v_z(x,z)&=&g(z) + \dfrac{h\,(z)}{\ln \xi^{-1}+B} 
+{\cal O}\left(\frac{1}{(\ln \xi)^2} \right).\label{scz}
\eee
The leading terms in the 
expansion of $z$ around $z_{eq}$ give  
$f(z)\!\approx\!-A$, $g(z)\!\approx\!-C(z\!-\!z_{eq})$ and 
$h(z)\!\approx\!-D$.
The constants
$A\!=\!0.80$, $C\!=\!0.124$ and $D\!=\!0.22$ 
are evaluated numerically
as $A\!=\!-(\partial v_x(x,z_{eq})/\partial x)_{x=1}$, 
$C\!=\!-(\partial 
v_z(1,z)/\partial z)_{z=z_{eq}}$ and 
$D=-\{\partial v_z(x,z_{eq})/\partial [1/(\ln \xi^{-1}+b)]\}_{x=1}$.
Asymptotic approximation of  Eqs.~\eqref{d1}-\eqref{d2}~reads
\bee
\dot{\xi}&=&-A\xi,\label{xclose}\\
\dot{z}&=&-C(z-z_{eq})-\dfrac{D}{\ln \xi^{-1}+B}. \label{zclose}
\eee
By integrating Eq.~\eqref{xclose} and the trajectory equation, 
\bee
\frac{d z}{d \xi}=\frac{C}{A\xi}(z-z_{eq})+\dfrac{D}{A\xi (\ln \xi^{-1}+B)}, 
\label{tra}
\eee
the asymptotic solution is easily obtained,
\bee
\xi&=&\xi_0\;e^{-At},\label{xiexp}\\
z\!-\!z_{eq}&=&\left\{(z_0\!-\!z_{eq})e^{\tau_0}\! - (D/A) 
[Ei(\tau)-Ei(\tau_0)]\right\} 
e^{-\tau}\!\!,\;\;\;\nonumber\\ \label{za}
\eee
where $\tau=C(\ln\xi^{-1}+B)/A=C[t+(\ln\xi_0^{-1}+B)/A]$, and its  
value at $t=0$ is equal to $\tau_0=C(\ln\xi_0^{-1}+B)/A$. As before,  
$\xi_0=x_0-1$.  The symbol 
$Ei(\tau)$ denotes the exponential integral,  
the Cauchy principal value of $\int_{-\infty}^{\tau}  dt
e^t/t$. 

Precision of the asymptotic solutions \eqref{xiexp}-\eqref{za} 
is illustrated in the inset of Fig.~\ref{zt}.
For gaps $\xi \lesssim 10^{-4}$, the numerical $z(t)$  
is approximated with  5\% accuracy by the  asymptotic one. Similar 
estimation holds for $\ln \xi(t)$. 

For very large times (for very small gaps $\xi$), 
$\tau \rightarrow \infty$, 
%the exponential integral 
and $Ei(\tau)\sim (1+1/\tau)\,e^{\tau}/\tau$. 
In this case, relation \eqref{za} 
further 
simplifies,
\bee
z-z_{eq} \sim -\frac{D}{C}\; \dfrac{1}{\ln \xi^{-1}+B} 
\hspace{1cm}\mbox{if }\xi \rightarrow 0,\;\;\;\label{asy}
\eee
and $(x,z) \rightarrow (x_{eq},z_{eq})$, if $t\rightarrow \infty$.
The equilibrium solution 
$(x_{eq},z_{eq})$ of the dynamics 
\eqref{d1}-\eqref{d2} is stable.

\begin{figure}
[th]
\psfrag{x13}{\large $\;\;\;\;\;\;x/2$}
\psfrag{z12}{\large $\!\!\!\!\!\!z\;$}
\psfrag{zz12}{ }
{\includegraphics[width=89mm,angle=0]{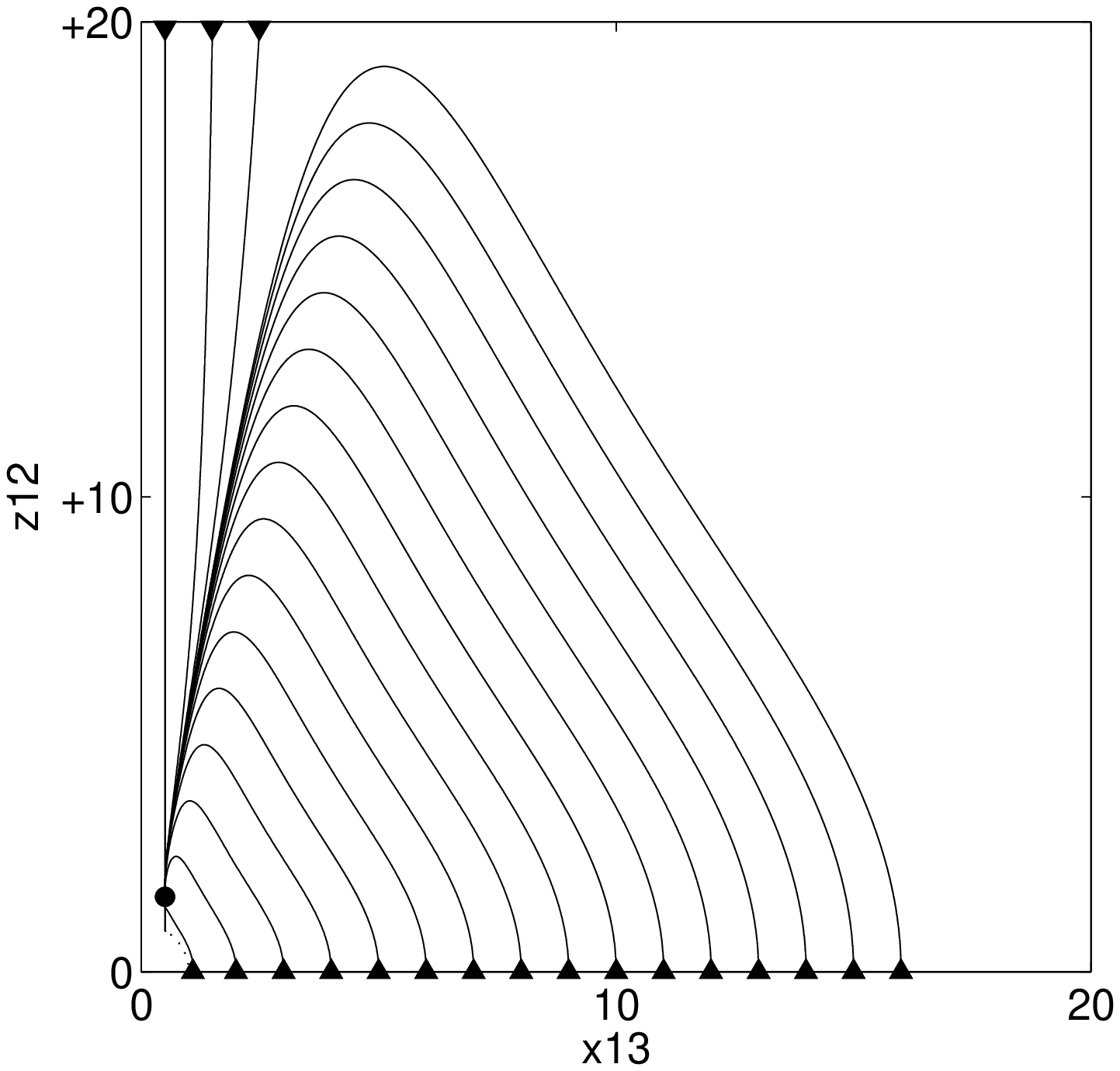}}
\begin{picture}(1, 1)(-27.5, -149)
\resizebox{3.3cm}{!}{\includegraphics{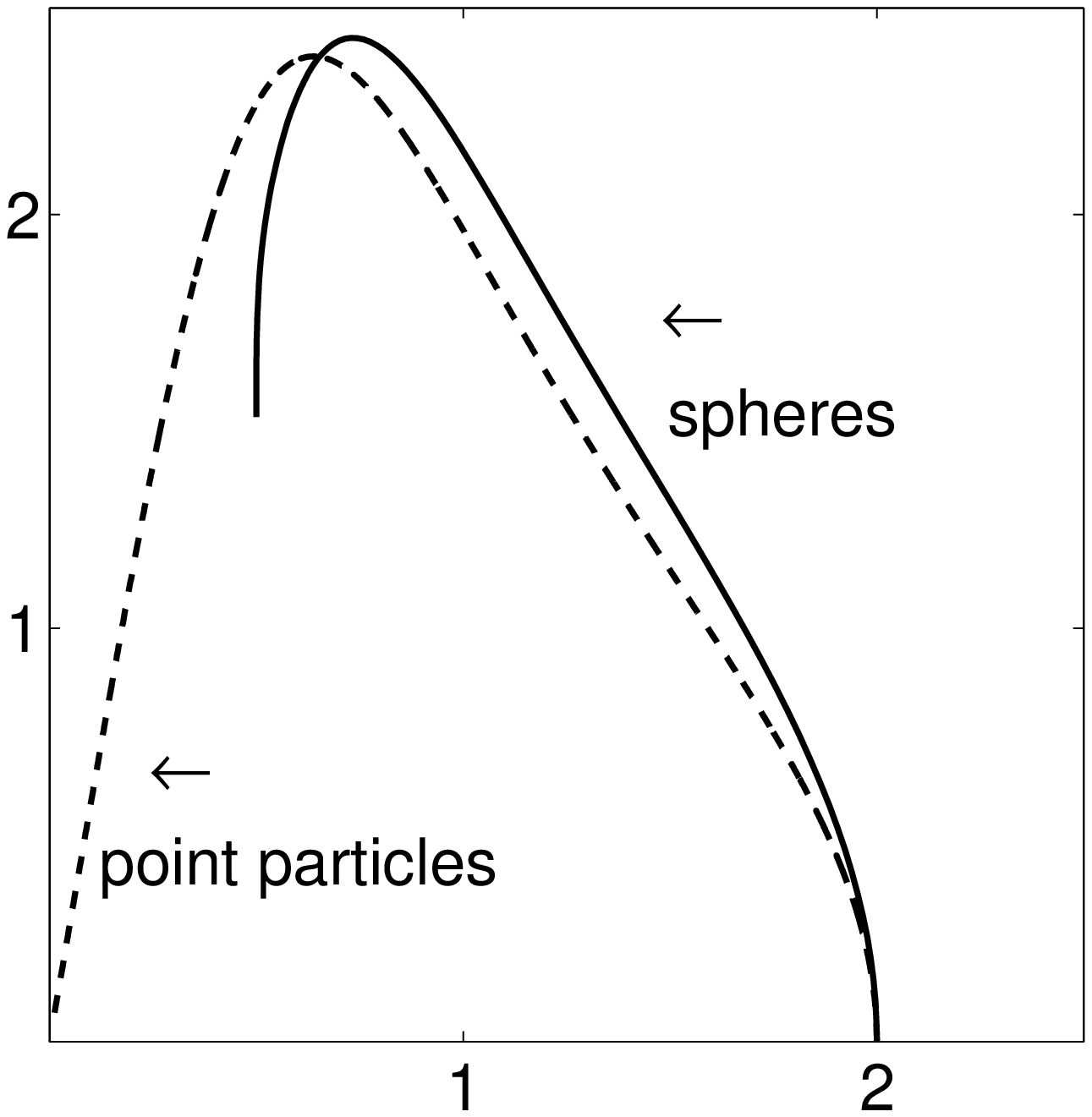}}
\end{picture}
\caption{Phase portrait of sphere dynamics \eqref{d1}-\eqref{d2}.
The stable node is denoted by  $\bullet$.
Inset: sphere trajectory (solid line) is compared with its point-particle 
counterpart
(dashed line).}\label{portrait}
\end{figure}

Finally, in Fig.~\ref{portrait}  we numerically evalute the phase portrait 
for the dynamics \eqref{d1}-\eqref{d2}. 
The two-dimensional 
phase space excludes non-overlapping spheres, and is therefore 
given as
$\{(x,z):\;x\ge 1 \mbox{ and } x^2/4 + z^2\ge 1\}$. 
The equilibrium
$(x_{eq},z_{eq})$ is a stable improper node. At this point, 
$dz/dx=- \infty$ for all the trajectories. Indeed, 
$dz/dx \sim -(D/C)/[(\ln \xi^{-1}+B)^2 \xi]$ close to  
$(x_{eq},z_{eq})$, as it follows from 
Eq.~\eqref{asy}. 
The trajectories of the spheres initially aligned horizontally 
keep the same shape as long as 
all the spheres are at large distances from each other.
As illustrated in the inset of Fig.~\ref{portrait}, 
this ``ideal'' shape is given by the scale-free 
point-particle dynamics.  
In Fig.~\ref{portrait}, another family of trajectories is also plotted, 
with $z=\infty$ at $t=-\infty$. These trajectories never reach $z=0$.

For $z<0$, trajectories and the motion are easily obtained by the time 
reversal superposed with the reflection in the plane $z=0$, using  
Eq. \eqref{sy2}. In particular, 
$(x_{eq},-z_{eq})$ is unstable improper node. 
Summarizing, the equilibrium points of the sphere dynamics 
\eqref{d1}-\eqref{d2} are located on 
the curve $z^2+x^2/4=1$ (marked by a dotted line in Fig.~\ref{portrait}) 
and at $(x_{eq},\pm z_{eq})$.
On the contrary,  no equilibrium exists for the 
relative motion of point-particles, as it follows from Eq.~\eqref{pp}.

To conclude, a  stable 
equilibrium has been found for a symmetric relative motion of three identical 
spheres settling under gravity, in which the sphere centers form an isosceles 
vertical
triangle with the horizontal base. In the equilibrium configuration, 
the two upper spheres touch each other and are  well-separated
from the lower singlet. For a large class of initial conditions, including 
distant spheres, the system 
evolves towards the equilibrium, reaching it after infinite time. 
The corresponding point-force approximation breaks down, because after 
a finite time, all the particles experience ``the end-of-world'', 
collapsing onto a single point. If two 
spheres are far below the single one, the system separates, increasing 
the relative vertical distance with time.  
In addition to the physical importance of the results, the solutions 
presented here may be used as benchmarks for numerical simulations of 
many-particle systems.

%\newpage

\end{document}